includes a few options for different layouts and
content for various journals. Please consult a previous issue of
your journal as needed.

\documentclass[smallextended]{svjour3}       
\usepackage{graphicx}
\usepackage{bbm}
\usepackage{mathptmx}      
\usepackage{latexsym}
\journalname{Quantum Information Processing}

\begin{document}

\title{Robust quantum spatial search}
\author{Avatar Tulsi}
\institute{Avatar Tulsi \at
              Department of Physics, IIT Bombay, Mumbai - 400076, India \\
              Tel.: +91-22-2576-7596\\
              \email{tulsi9@gmail.com}}

\date{Received: date / Accepted: date}

\maketitle

\begin{abstract}
Quantum spatial search has been widely studied with most of the study focusing on quantum walk algorithms. We show that quantum walk algorithms are extremely sensitive to systematic errors. We present a recursive algorithm which offers significant robustness to certain systematic errors. To search $N$ items, our recursive algorithm can tolerate errors of size $O(1/\sqrt{\ln N})$ which is exponentially better than quantum walk algorithms for which tolerable error size is only $O(\ln N/\sqrt{N})$. Also, our algorithm does not need any ancilla qubit. Thus our algorithm is much easier to implement experimentally compared to quantum walk algorithms.
\end{abstract}

\section{Introduction}
\label{intro}

An important application of quantum algorithms is quantum spatial search (QSS) where we search a database of items spatially distributed on the vertices of an underlying lattice. The locality constraint must be satisfied which demands that in one time step, we can implement only one local operator coupling only neighboring vertices. The famous Grover's algorithm is optimal for general search problems~\cite{grover,qaa1,qaa2,optimal} but it becomes inefficient for QSS. However, it can be generalized to get efficient QSS algorithms. 

	First efficient QSS algorithm was a recursive application of Grover's algorithm~\cite{recursive} but subsequent algorithms were based on quantum walk (QW)~\cite{akr,cg1,cg2}. QW algorithms have been widely investigated both theoretically and experimentally as they offer better time complexity. Unlike their construction and performance, their sensitivity to errors has not been much explored. Errors are inevitable and pose the biggest challenge in experimental implementation of algorithms. Completely random errors can only be handled using the methods of quantum error correction (QEC) and fault-tolerant quantum computation (FTQC). These methods use extra expensive resources to add redundancy to the quantum states and gates to overcome small errors. For systematic errors exhibiting specific structures, it is important to investigate the possibility of quantum algorithms which are intrinsically robust to such errors so that we don't need expensive resources used by QEC and FTQC.

	In this letter, first we show that QW algorithms are not robust as they are extremely sensitive to systematic errors. Then we present a recursive algorithm which is intrinsically robust to such errors. We assume systematic errors to be reproducible and reversible. Suppose we try to implement an operator $\mathcal{Y}$ but due to errors, what we actually implement is $\mathcal{Z}$. Error reproducibility means that every time we try to implement $\mathcal{Y}$, we actually implement $\mathcal{Z}$. Error reversibility means that every time we try to implement $\mathcal{Y}^{\dagger}$, we actually implement $\mathcal{Z}^{\dagger}$. Such errors are not uncommon and they arise when there is incorrect calibration of the instrumentation. For example, they arise due to imperfect pulse calibration and offset effect in NMR systems~\cite{nmr}. 

	We only consider the case of two-dimensional square lattice as similar ideas can be used for other lattice structures and higher dimensions. Our algorithm is based on recursion. Unlike the previous recursive algorithm~\cite{recursive}, our algorithm does not need ancilla qubits to work. Ancilla qubits are a hurdle in the experimental implementation and it is preferable to avoid their usage. The original versions of all quantum walk (QW) algorithms needed ancilla qubits either to implement coinspace in DTQW or spin degrees of freedom in CTQW. But subsequently, it has been shown that QW algorithms do not need any ancilla qubit~\cite{graphene1,graphene2,childscoin,patelcoin,patel1,patel2,falk,andris}, a notable work being the CTQW algorithm for QSS on a graphene lattice~\cite{graphene1} which is also experimentally implemented in ~\cite{graphene3}. We extend this advantage to recursive algorithms and thus we show that recursive algorithms are easier to implement experimentally.	

\section{Robustness of QW algorithms}
\label{QWrobustness}

	Let us first consider the discrete time quantum walk (DTQW) algorithm for QSS presented by Ambainis, Kempe and Rivosh (AKR)~\cite{akr}. To analyze its robustness, we consider it as a special case of the general quantum search algorithm (GQSA) which iterates $\mathcal{S} = D_{s}I_{t}^{\phi}$ on $|s\rangle$ to take it close to $|t\rangle$~\cite{general}. Here $I_{t}^{\phi}$ is the selective phase rotation of $|t\rangle$ and $D_{s}$ can be any unitary operator satisfying $D_{s}|s\rangle = |s\rangle$. 

	Let $D_{s}|\ell\rangle = e^{\imath \theta_{\ell}}|\ell\rangle$ be the eigenspectra of $D_{s}$ with $\theta_{\ell} \in [-\pi,\pi]$ and $\theta_{\ell = s} = 0$. Let $\theta_{\rm min} \leq |\theta_{\ell \neq s}|$ be the spectral gap and $e^{\imath \lambda_{\pm}}$ be the two eigenvalues of $\mathcal{S}$ satisfying $|\lambda_{\pm}| \leq \theta_{\rm min}$. Assuming $|\langle s|t\rangle| \ll 1$ and $|\lambda_{\pm}|\ll \theta_{\rm min}$, the performance of GQSA is determined by the moments, 
\begin{equation}
\Lambda_{p} = \sum_{\ell \neq s}|\langle \ell |t\rangle|^{2}\cot^{p}\left(\theta_{\ell}/2\right), \ \ \ p \in \{1,2\}. 
\end{equation}
Eq. (27) of ~\cite{general} implies that 
\begin{equation}
|\langle t|\mathcal{S}^{q_{\rm m}}|s\rangle| = \sqrt{P_{\rm m}} = \frac{\sin 2\eta}{B\sin (\phi/2)},\ q_{\rm m} = \frac{\pi B\sin 2\eta}{4|\langle t|s\rangle|},
\end{equation}
where $\eta \in [0,\pi/2]$ is determined by
\begin{equation}
\cot 2\eta = \frac{A}{2B|\langle t|s\rangle|},\ A = \cot\frac{\phi}{2} + \Lambda_{1},\ B^{2} = 1+\Lambda_{2}.
\end{equation} 
The state $\mathcal{S}^{q_{\rm m}}|s\rangle$ can be evolved to $|t\rangle$ using Grover's algorithm by applying $\Theta(1/\sqrt{P_{\rm m}})$ iterations of $\mathcal{S}^{q_{\rm m}}I_{s}\left(\mathcal{S}^{q_{\rm m}}\right)^{\dagger}I_{t}^{\pi}$. Thus the time complexity of GQSA is
\begin{equation}
T_{GQS} = \Theta\left(P_{\rm m}^{-1/2}(2q_{\rm m}T[\mathcal{S}] + T[I_{s}]+T[I_{t}^{\pi}])\right),
\end{equation}   
where $T[X]$ is the number of time steps needed to implement $X$. Thus $T[X]$ is $T[X^{\dagger}]$, and $T[XX']$ is $T[X] + T[X']$. 

	For AKR algorithm, $T[L]$ is $1$ for any local operator $L$ which includes $I_{t}^{\phi}$ and $D_{s}$. But $I_{s}$ is a nonlocal operator coupling each vertex to all other vertices and $T[I_{s}]$ is $2\sqrt{N}$ as shown in ~\cite{akr}. Also, $|\langle t|s\rangle|$ is $1/\sqrt{N}$ and  
\begin{equation}
T_{AKR} = \sqrt{N}B\sin(\phi/2)\left[(\pi B/2) + \csc 2\eta\right] \label{timeAKR}
\end{equation}
In case of no errors, $\phi$ is $\pi$ and $D_{s}$ is a real orthogonal operator for which $\Lambda_{1}$ is $0$ and $\Lambda_{2}$ is $\Theta(\ln N)$. Thus $A$ is $0$, $B$ is $\Theta(\sqrt{\ln N})$ and $T_{AKR}$ is $\Theta(\sqrt{N}\ln N)$. The systematic errors cause disturbance by shifting $A$ from its ideal value of $0$, i.e. $A \neq 0$. For AKR algorithm, $\cot 2\eta$ is $A\Theta(\sqrt{N/\ln N})$. If $A \gg \ln N/\sqrt{N}$ then $\csc 2\eta \approx A\Theta(\sqrt{N/\ln N}) \gg \pi B/2$. Using Eq. (\ref{timeAKR}), we get $T_{AKR} = AN \Theta(\sin\frac{\phi}{2})$. 

	A main reason for $A \neq 0$ is systematic phase errors (SPE) which shift $\phi$ from $\pi$ to $\pi + \epsilon$. For small errors, $\epsilon \ll 1$ and $A$ is $\epsilon/2$ assuming $\Lambda_{1} =0$. For $\epsilon = 2A = \Omega(\ln N/\sqrt{N})$, $T_{AKR}$ is $\Theta(\epsilon)N$, much larger than the optimal $\Theta(\sqrt{N})$ complexity. Typically, $N \gg 1$ and $\ln N/\sqrt{N} \ll 1$ hence AKR algorithm is extremely sensitive to SPE. This proves analytically the numerical observation of ~\cite{numerical}. It is similar to the sensitivity of search algorithms to SPE due to phase-matching condition~\cite{gaterandom,phasematching1,phasematching2} which is a source of dominant gate imperfection and poses an intrinsic limitation to the size of database $N$ that can be searched~\cite{dominantgate}.

	A non-zero $\Lambda_{1}$ also implies $A \neq 0$. Ideally, $D_{s}$ is a real orthogonal operator and its eigenphase distribution has a symmetry which causes $\Lambda_{1} = 0$. The existence of $\theta_{\ell}$ implies the existence of $\theta_{\ell'} = -\theta_{\ell}$ such that $|\langle t|\theta_{\ell}\rangle| = |\langle t|\theta_{\ell'}\rangle|$. Even a slight violation of this symmetry can imply $|A| = |\Lambda_{1}| \gg \ln N/\sqrt{N}$ and cause AKR algorithm to fail. An example was presented in ~\cite{thomas1} where $D_{s}$ is not a perfect real operator. If we know $\Lambda_{1}$ then we can choose $\phi = \phi'$ such that $A = \cot\frac{\phi'}{2} + \Lambda_{1} = 0$ and AKR algorithm works. For the example of ~\cite{thomas1}, this was shown in ~\cite{thomas2}. But $\phi$ can deviate from $\phi'$ due to SPE. Thus SPE is a dominant source of error.   

	For continuous time quantum walk (CTQW) algorithms, the search is an evolution under the Hamiltonian $H = \gamma \mathcal{L} + |t\rangle\langle t|$, where $\mathcal{L}$ is the Laplacian of underlying graph of vertices. CTQW algorithms are extremely sensitive to the value of $\gamma$ and fail for a square lattice if $|\gamma - \gamma_{c}| \gg \ln N/\sqrt{N}$ where $\gamma_{c}$ is the critical value of $\gamma$. This is equivalent to the sensitivity of DTQW algorithms to the value of $\phi$. Thus QW algorithms are not robust. Next, we present a robust algorithm for QSS.

\section{Preliminaries}
\label{preliminaries}	

	Our search space is a $\sqrt{N} \times \sqrt{N}$ lattice. We assume $\sqrt{N} = 3^{n}$ for integer $n$. For $\kappa \in \{1,2,\ldots,n\}$, we define 
\begin{equation}
\mathbbm{N}\kappa = \{0,1,\ldots,3^{\kappa}-1\},\ \ \ \bar{\kappa}= n-\kappa.
\end{equation} 
We label $N$ lattice vertices by their $x$ and $y$ coordinates where $(x,y) \in \mathbbm{N}n$. In the quantum scenario, the vertices $(x,y)$ are encoded by the basis states $|x,y\rangle$ of a $N = 3^{2n}$-dimensional Hilbert space $\mathcal{H}_{n}$. For any $\kappa$, $\mathcal{H}_{n}$ can be partitioned into $3^{2\bar{\kappa}} = 9^{n-\kappa}$ subspaces $\mathcal{H}^{\alpha\beta}_{\kappa}$ where $(\alpha,\beta) \in \mathbbm{N}\bar{\kappa}$ with each subspace encoding a $3^{\kappa} \times 3^{\kappa}$ subsquare. Explicitly, $\mathcal{H}^{\alpha\beta}_{\kappa}$ is spanned by the basis states
\begin{equation}
|x,y\rangle_{\kappa}^{\alpha\beta}  \equiv |3^{\kappa}\alpha + x, 3^{\kappa}\beta + y\rangle,\ \  (x,y) \in \mathbbm{N}\kappa. \label{uvdefine}
\end{equation} 
For each subsquare, let 
\begin{equation}
|s_{\kappa}^{\alpha \beta}\rangle = 3^{-\kappa}\sum_{x,y \in \mathbbm{N}\kappa}|x,y\rangle_{\kappa}^{\alpha\beta}. \label{uniformstate}
\end{equation}
be the uniform superposition state (\emph{u.s.s.}) of all subsquare vertices.

	For any quantum state $|\chi\rangle$ and angle $\vartheta$, let $I^{\vartheta}(\chi)$ be the selective phase rotation by angle $\pi+\vartheta$ of $|\chi\rangle$, i.e.
\begin{equation}
I^{\vartheta}(\chi) = \mathbbm{1}-f_{\vartheta}|\chi\rangle\langle \chi|,\ \ f_{\vartheta} = 1-e^{\imath (\pi + \vartheta)} = 1+e^{\imath \vartheta}. \label{selectivephaserotationdefine}
\end{equation} 
If $|\chi\rangle = V|\chi_{0}\rangle$ for a unitary operator $V$ then 
\begin{equation}
I^{\vartheta}(\chi) = 1-f_{\vartheta}V|\chi_{0}\rangle \langle \chi_{0}|V^{\dagger} = VI^{\vartheta}(\chi_{0})V^{\dagger}. \label{selectivedefine2}
\end{equation} 
For $\vartheta = 0$, $I^{0}(\chi)$ is the selective inversion of $|\chi\rangle$. Here $\vartheta$ is basically an error parameter. The desired operator is $I^{0}(\chi)$ corresponding to zero error but due to errors, available operator is $I^{\vartheta}(\chi)$.  

	We define the operator $\mathcal{I}^{\delta}(s_{\kappa})$ as  
\begin{equation}
\mathcal{I}^{\delta}(s_{\kappa}) =  \prod_{\alpha,\beta \in \mathbbm{N}\bar{\kappa}}I^{\delta}(s_{\kappa}^{\alpha\beta}) = S_{\kappa}I^{\delta}(00\kappa)S_{\kappa}^{\dagger}. 
\label{Isdefine}
\end{equation}
Here $S_{\kappa}$ satisfy $S_{\kappa}|0,0\rangle_{\kappa}^{\alpha\beta} = |s_{\kappa}^{\alpha\beta}\rangle$ for all $(\alpha,\beta) \in \mathbbm{N}\bar{\kappa}$ and $I^{\delta}(00\kappa) = \prod_{\alpha,\beta \in \mathbbm{N}\bar{\kappa}}I^{\delta}(|0,0\rangle_{\kappa}^{\alpha\beta})$. We have
\begin{equation}
T[I^{\delta}(s_{\kappa})] = 2T[S_{\kappa}] + 1, \label{stimesteps1}
\end{equation}
where $T[X]$ is defined earlier and $T[I^{\delta}(00\kappa)] = 1$ as $I^{\delta}(00\kappa)$ is a local operator. 

	To find $T[S_{\kappa}]$, we present a method to implement $S_{\kappa}$. Suppose the initial state is $|0,y\rangle_{\kappa}^{\alpha\beta}$ for any $y$. We define the states $|\mu\rangle_{a}$ for $a \in \mathbbm{N}\kappa$ as
\begin{equation}
\sqrt{3^{\kappa}}|\mu\rangle_{a} = \sum_{x=0}^{a-1}|x,y\rangle_{\kappa}^{\alpha\beta} + \sqrt{3^{\kappa}-a}|a,y\rangle_{\kappa}^{\alpha\beta}. \label{phidefine}
\end{equation} 
For $a=0$, we choose $\sum_{x=0}^{a-1}|x,y\rangle_{\kappa}^{\alpha\beta}$ to be a null vector. So $|\mu\rangle_{0}$ is the initial state 
$|0,y\rangle_{\kappa}^{\alpha\beta}$.	We define local operators $L_{b}$ for $b \in \mathbbm{N}\kappa$ as
\begin{equation}
\sqrt{3_{b}^{\kappa}}L_{b}|b-1,y\rangle_{\kappa}^{\alpha\beta}  =  |b-1,y\rangle_{\kappa}^{\alpha\beta} + \sqrt{3_{b}^{\kappa}-1}|b,y\rangle_{\kappa}^{\alpha\beta}, \label{Ldefine}
\end{equation}
where $3_{b}^{\kappa} = 3^{\kappa}-b+1$. Using eqs. (\ref{phidefine},\ref{Ldefine}) and little calculation, we get $L_{b}|\mu\rangle_{b-1} = |\mu\rangle_{b}$. We define $S_{\kappa,x}$ as
\begin{equation}
S_{\kappa,x}|0,y\rangle_{\kappa}^{\alpha\beta} = \prod_{x \in \mathbbm{N}\kappa}L_{b}|\mu\rangle_{0} = |\mu\rangle_{3^{\kappa}-1}. \label{Sxdefine}
\end{equation}
As $T[L_{b}] = 1$, we have $T[S_{\kappa,x}] = 3^{\kappa}-1$. It is easy to check that $|\mu\rangle_{3^{\kappa}-1} \propto \sum_{x \in \mathbbm{N}\kappa}|x,y\rangle_{\kappa}^{\alpha\beta}$ is the normalized state of 
uniform distribution of amplitudes in $x$-direction over the vertices of $\mathcal{H}^{\alpha\beta}_{\kappa}$. The roles of $x$ and $y$ directions can be interchanged to design an operator $S_{\kappa,y}$ such that $T[S_{\kappa,y}] = T[S_{\kappa,x}] = 3^{\kappa}-1$ and
\begin{equation}
S_{\kappa,y}|x,0\rangle_{\kappa}^{\alpha\beta} = 3^{-\kappa/2}\sum_{y \in \mathbbm{N}\kappa}|x,y\rangle_{\kappa}^{\alpha\beta}, \label{Sydefine}
\end{equation} 
for any value of $x$. Eqs. (\ref{Sxdefine}) and (\ref{Sydefine}) imply that the operator $S_{\kappa} = S_{\kappa,x}S_{\kappa,y}$ satisfies $S_{\kappa}|0,0\rangle_{\kappa}^{\alpha\beta} = |s_{\kappa}^{\alpha\beta}\rangle$. Thus $T[S_{\kappa}] = 2T[S_{\kappa,x}] = 2(3^{\kappa}-1)$ and Eq. (\ref{stimesteps1}) implies
\begin{equation}
T[S_{\kappa}] = 2(3^{\kappa}-1) \Longrightarrow T[\mathcal{I}^{\delta}(s_{\kappa})] = 4\cdot3^{\kappa} - 3. \label{stimesteps}
\end{equation}

\section{Algorithm}
\label{algorithm}
	
Consider the recursive relation, 
\begin{equation}
|\psi_{\kappa}\rangle = U_{\kappa}|t\rangle =  I^{\epsilon}(\psi_{\kappa-1})\mathcal{I}^{\delta}(s_{\kappa})|\psi_{\kappa-1}\rangle,\ \ \ |\psi_{0}\rangle = |t\rangle. \label{recursion}
\end{equation} 
Eq. (\ref{selectivedefine2}) yields $I^{\epsilon}(\psi_{\kappa}) =  U_{\kappa} I^{\epsilon}(t)U_{\kappa}^{\dagger}$. Thus 
\begin{equation}
U_{\kappa} = U_{\kappa-1} I^{\epsilon}(t)U_{\kappa-1}^{\dagger} \mathcal{I}^{\delta}(s_{\kappa})U_{\kappa-1},\ \ U_{0} = \mathbbm{1}_{N}.  \label{recursionforU}
\end{equation}
Eq. (\ref{stimesteps}), with $T[I^{\epsilon}(t)] = 1$ as $I^{\epsilon}(t)$ is local, implies 
\begin{eqnarray}
T[U_{\kappa}] &=& 3T[U_{\kappa -1}] + T[\mathcal{I}^{\delta}(s_{\kappa})]  + T[I^{\epsilon}(t)] \nonumber \\
 & = & 3T[U_{\kappa -1}] + 4\cdot 3^{\kappa} - 2, \ \ T[U_{0}] = 0.
\end{eqnarray}
Solving above recursive relation, we get
\begin{equation}
T[U_{\kappa}] = (4\kappa-1) 3^{\kappa} + 1.  \label{solverecursive}
\end{equation}

	We note that $|t\rangle$ is a unique vertex. For any $\kappa$, let $\alpha\beta = \alpha_{t}\beta_{t} = \tau$ be the index of subspace $\mathcal{H}_{\kappa}^{\tau}$ containing the target state $|t\rangle$ and let $\mathcal{H}_{\kappa}^{\neq \tau}$ be its complementary subspace. Let $|x,y\rangle_{\kappa}^{\tau}$ denote the $9^{\kappa}$ basis states of $\mathcal{H}_{\kappa}^{\tau}$ for $(x,y) \in \mathbbm{N}\kappa$ and let $|x,y\rangle_{\kappa}^{\neq \tau}$ denote the $9^{n} - 9^{\kappa}$ basis states of $\mathcal{H}_{\kappa}^{\neq \tau}$. Let $|s_{\kappa}^{\tau}\rangle$ be the $\emph{u.s.s.}$ of $|x,y\rangle_{\kappa}^{\tau}$, i.e.
\begin{equation}
|s_{\kappa}^{\tau}\rangle = 3^{-\kappa}\sum_{x,y \in \mathbbm{N}\kappa}|x,y\rangle_{\kappa}^{\tau}.  \label{skgamma}
\end{equation} 
By definition, $\mathcal{H}_{\kappa}^{\tau} \subset \mathcal{H}_{\kappa + 1}^{\tau}$ and $\mathcal{H}_{\kappa + 1}^{\neq \tau} \subset \mathcal{H}_{\kappa}^{\neq \tau}$. Then Eq. (\ref{skgamma}) implies that the component of $|s_{\kappa+1}^{\tau}\rangle$ in each basis state $|x,y\rangle_{\kappa}^{\tau}$ is $3^{-\kappa-1}$. Hence $\langle s_{\kappa +1}^{\tau}|s_{\kappa}^{\tau}\rangle = 1/3$ or
\begin{equation}
|s_{\kappa + 1}^{\tau}\rangle = (1/3)|s_{\kappa}^{\tau}\rangle + \sum h_{x,y}|x,y\rangle_{\kappa}^{\neq\tau} \label{kappaplus1expand} 
\end{equation}
We define $\alpha_{\kappa} = \langle s_{\kappa}^{\tau} | \psi_{\kappa -1} \rangle$ and assume 
\begin{equation}
\langle \psi_{\kappa -1}| x,y\rangle_{\kappa}^{\neq \tau} = 0 \label{recursiveassume}
\end{equation} 
for all basis states $|x,y\rangle_{\kappa}^{\neq \tau}$. Then $I^{\delta}(s_{\kappa}^{\alpha\beta \neq \tau})$ leave $|\psi_{\kappa-1}\rangle$ unchanged and $\mathcal{I}^{\delta}(s_{\kappa}) \equiv I^{\delta}(s_{\kappa}^{\tau})$ for $|\psi_{\kappa -1}\rangle$. Thus
\begin{equation}
|\psi_{\kappa}\rangle \equiv I^{\epsilon}(\psi_{\kappa-1})I^{\delta}(s_{\kappa}^{\tau})|\psi_{\kappa-1}\rangle, \label{recursion2}
\end{equation} 
using Eq. (\ref{recursion}). Then Eq. (\ref{selectivephaserotationdefine}) and little calculation imply
\begin{equation}
|\psi_{\kappa}\rangle = \left(1-f_{\epsilon} + f_{\epsilon}f_{\delta}|\alpha_{\kappa}|^{2}\right)|\psi_{\kappa-1}\rangle - \alpha_{\kappa}f_{\delta}|s_{\kappa}^{\tau}\rangle. \label{kappaplus1}
\end{equation}
As $\mathcal{H}_{\kappa + 1}^{\neq \tau} \subset \mathcal{H}_{\kappa}^{\neq \tau}$, Eq. (\ref{recursiveassume}) implies $\langle \psi_{\kappa -1}| x,y\rangle_{\kappa+1}^{\neq \tau} = 0$. By definition, $\langle s_{\kappa}^{\tau}| x,y\rangle_{\kappa+1}^{\neq \tau} = 0$ and Eq. (\ref{kappaplus1}) implies that Eq. (\ref{recursiveassume}) is true for $\kappa + 1$ if it is true for any $\kappa$. It is true for $\kappa = 1$ as $|\psi_{0}\rangle = |t\rangle$ is orthogonal to $\mathcal{H}_{1}^{\neq \tau}$. Hence eq. (\ref{recursiveassume}) is true for all $\kappa$. Eq. (\ref{kappaplus1expand}) implies an immediate consequence $\langle s_{\kappa + 1}^{\tau}|\psi_{\kappa - 1}\rangle = \alpha_{\kappa}/3$ putting which, along with $\langle s_{\kappa + 1}^{\tau}|s_{\kappa}^{\tau}\rangle = 1/3$, in eq. (\ref{kappaplus1}), we get 
\begin{equation}
\alpha_{\kappa + 1} = (\alpha_{\kappa}/3)\left[1-f_{\epsilon}-f_{\delta} + f_{\epsilon}f_{\delta}|\alpha_{\kappa}|^{2}\right]. \label{recursionomega1}
\end{equation}
as the recursive relation for $\alpha_{\kappa}$.

	For small errors, we have $\Delta = \max(|\delta|,|\epsilon|) \ll 1$ and $f_{\vartheta} \approx 2-(\vartheta^{2}/2) + \imath \vartheta$ for $\vartheta \in \{\delta, \epsilon\}$. Retaining only leading second order terms in Eq. (\ref{recursionomega1}), we get the following relation for $\omega_{\kappa} = |\alpha_{\kappa}|^{2}$, 
\begin{equation}
\frac{\omega_{\kappa+1}}{\omega_{\kappa}} = \left(1-\frac{4\omega_{\kappa}}{3}\right)^{2} - \frac{\tilde{\omega}_{\kappa}}{9}\left[\epsilon^{2}+\delta^{2} + \tilde{\omega}_{\kappa}(\epsilon - \delta)^{2} \right], \label{recursionomegakappa}
\end{equation}
where $\tilde{\omega}_{\kappa} = 1-2\omega_{\kappa}$. For $\omega_{\kappa} < 1/2$, $\tilde{\omega}_{\kappa} > 0$ and $1-(4\omega_{\kappa}/3)$ is $\Theta(1)$. Hence $\omega_{\kappa} >\omega_{\kappa + 1} = \Theta(\omega_{\kappa})$. As $|\psi_{0}\rangle = |t\rangle$, $\alpha_{1} = \langle s_{1}^{\tau}|\psi_{0}\rangle = 1/3$ and $\omega_{1} = 1/9 < 1/2$. Thus, as $\kappa$ increases, $\omega_{\kappa}$ decreases by a constant factor. Suppose, for $\kappa = \kappa0$, $\omega_{\kappa0} = 1/10n$. If $n \leq \kappa0$ then $\omega_{n} \geq 1/10n$ else Eq. (\ref{recursionomegakappa}) implies
\begin{eqnarray}
\omega_{n}  & \geq & (1/10n)[1-(4n/15)-(2\Delta^{2}/3)]^{n-\kappa0} \nonumber \\
   & \geq & 11/150n - \Delta^{2}/15,
\end{eqnarray} 
retaining only leading order terms and assuming $\Delta \ll 1/\sqrt{n} = 1/\sqrt{\ln N}$. Thus $\omega_{n} = |\alpha_{n}|^{2} = \Omega(1/\ln N)$ as long as errors are small. Using Eq. (\ref{recursion}) and $\alpha_{\kappa} = \langle s_{\kappa}^{\tau} | \psi_{\kappa -1} \rangle$, we get $|\langle s_{n}|U_{n-1}|t\rangle| = |\langle t|U_{n-1}^{\dagger}|s_{n}\rangle| = \Omega(1/\sqrt{\ln N})$. This amplitude can be amplified by using generalized quantum amplitude amplification with arbitrary phases $\delta$ and $\epsilon$ analyzed in~\cite{phasematching2}. As long as $|\epsilon - \delta| \ll 1/\sqrt{\ln N}$ (which is true as we assume $\Delta \ll 1/\sqrt{\ln N}$), the amplitude can be amplified to $\Theta(1)$ by $\Theta(\sqrt{\ln N})$ iterations of $U_{n-1}^{\dagger}I^{\delta}(s_{n})U_{n-1}I^{\epsilon}(t)$ on $U_{n-1}^{\dagger}|s_{n}\rangle$.  Thus the time complexity $Q$ of our algorithm is $\sqrt{\ln N}T[U_{n-1}]$. Putting $\kappa = n-1$ in Eq. (\ref{solverecursive}), we get $T[U_{n-1}] = \Theta{\sqrt{N} \ln N}$ and $Q = \sqrt{N} \ln^{3/2}N$.

\section{Robustness}
\label{robustness}
	
Our algorithm works as long as $\{\delta,\epsilon\} = O(1/\sqrt{\ln N})$. This tolerance to SPE is exponentially better than previous QSS algorithms which fail if $\epsilon \neq O(\ln N/\sqrt{N})$. Due to the term $U_{\kappa -1}^{\dagger}$ in Eq. (\ref{recursionforU}), our algorithm does not just need $\{I^{\delta}(s_{\kappa}),I^{\epsilon}(t)\}$ but also their reverse transformations $\{I^{-\delta}(s_{\kappa}),I^{-\epsilon}(t)\}$ which are available as errors are reversible.

	Our algorithm is also robust to local systematic errors in operators $S_{\kappa}$. If reversible errors perturb $S_{\kappa}$ to $\mathcal{E}^{\dagger}S_{\kappa}$ then it also perturbs $S_{\kappa}^{\dagger}$ to $S_{\kappa}^{\dagger}\mathcal{E}$. We have assumed $\mathcal{E}$ to be $\kappa$-independent for simplicity. As $I^{\delta}(s_{\kappa}) = S_{\kappa}I^{\delta}(00\kappa)S_{\kappa}^{\dagger}$, eq. (\ref{recursion}) becomes
\begin{eqnarray}
|\psi_{\kappa}\rangle & = & I^{\epsilon}(\psi_{\kappa-1})S_{\kappa}I^{\delta}(00\kappa)S_{\kappa}^{\dagger}|\psi_{\kappa-1}\rangle  \nonumber \\ 
  & = & I^{\epsilon}(\psi_{\kappa-1})\mathcal{E}^{\dagger}I^{\delta}(s_{\kappa})\mathcal{E}|\psi_{\kappa-1}\rangle. \label{recursionSkappa2}
\end{eqnarray}
Defining $|\psi_{k-1}'\rangle = \mathcal{E}|\psi_{k-1}\rangle$, above equation becomes $|\psi_{\kappa}'\rangle = I^{\epsilon}(\psi_{\kappa-1}')I^{\delta}(s_{\kappa})|\psi_{\kappa-1}'\rangle$ which is similar to Eq. (\ref{recursion}) and our analysis holds true provided assumption (\ref{recursiveassume}) is true. By inductive hypothesis, this is true when $|\psi_{0}'\rangle =  = \mathcal{E}|t\rangle$ is orthogonal to $\mathcal{H}_{1}^{\neq \tau}$. Assuming $|t\rangle$ to be the centre vertex of $3 \times 3$ subsquare $\mathcal{H}_{1}^{\tau}$, $\mathcal{E}|t\rangle$ is orthogonal to $\mathcal{H}_{1}^{\neq \tau}$ for all local errors $\mathcal{E}$. The ideal operator $S_{\kappa}$ is made up of local operators $L_{b}$ coupling $|x,y\rangle$ either to $|x+1,y\rangle$ or $|x,y+1\rangle$. It is reasonable to assume the same for errors affecting $L_{b}$ and hence locality of $\mathcal{E}$ is close to real situations. The general case of $\kappa$-dependendent $\mathcal{E}$ is hard to analyze and beyond the scope of this study.

\section{Discussion}
\label{discussion}

	Our algorithm takes $O(\sqrt{N}\ln^{3/2}N)$ time steps to search a square lattice. The best known performance is $O(\sqrt{N\ln N})$~\cite{fastersearch,fastergeneral,classicalambainis,postprocessing} but we need ancilla qubits for this. Without ancilla qubits, the best known performance is $O(\sqrt{N}\ln N)$. The slightly inferior time complexity could be a probable reason why recursive approach to QSS has not been as extensively studied as quantum walk based algorithms.   

	We have shown that recursive algorithms derive their importance due to significant robustness to systematic errors. Our algorithm works as long as errors are of the order of $1/\sqrt{\ln N}$ and hence error tolerance is exponentially better than that of QW algorithms which is $\ln N/\sqrt{N}$. Also, our algorithm works without any ancilla qubit and hence it is easier to implement. These observations indicate that our recursive algorithm for quantum spatial search is more likely to be implementable in near future. Thus it deserves a detailed theoretical and experimental investigation in the same way as QW based algorithms are currently being investigated.

\end{document}